\documentclass{iopart}
\usepackage{graphicx}
\usepackage{amssymb}
\usepackage{iopams}
\usepackage{setstack}
\usepackage{dsfont}

\begin{document}

\newcommand{\bra}[1]    {\left\langle #1\right|}
\newcommand{\ket}[1]    {| #1 \rangle}
\newcommand{\trc}[1]    {{\rm Tr}\left[ #1 \right]}
\newcommand{\av}[1]    {\langle #1 \rangle}
\newcommand{\x}{{\bf r}}
\newcommand{\proj}[1]{\ket{#1}\!\!\bra{#1}}

\title{Criteria for particle entanglement in many-body systems of bosons}

\author{T. Wasak, P. Sza\'nkowski, M. Trippenbach and J.~Chwede\'nczuk}
\address{Faculty of Physics, University of Warsaw, ul. Pasteura 5, PL-02-093 Warszawa, Poland}

\begin{abstract}
  Basing on the analogy between the coherent states of light and separable states of $N$ bosons, 
  we demonstrate that the violation Cauchy-Schwarz inequality 
  for any-order correlation function signals  the entanglement among the constituent particles. Rather than restricting to the correlations between the positions of particles, we
  consider the broadest set of measurements allowed by quantum mechanics.
  Our result is general -- it applies to any quantum system  of bosons, even when the number of particles is not fixed, provided that there is no coherence between different number states.
  We also demonstrate that the compact expression for the separable state of bosons can be used to relate some known metrological quantities to the particle entanglement in a very simple way.
\end{abstract}

\maketitle

\section{Introduction}

Although the foundations of quantum and classical physics are much different, it is often difficult to construct a simple criterion of non-classicality of a particular system.
A prominent example of a quantum phenomenon is the ability of particles to exist in superpositions of states.
The most well known manifestation of such superposition is the Young double-slit experiment for massive particles, which confirms their wave character and the ability to coherently interfere.
A more subtle example of genuinely quantum phenomenon is the Einstein-Podolsky-Rosen (EPR) paradox \cite{epr}, which was introduced using 
a pair of entangled particles, separated far apart. A consequence of quantum mechanics is that a measurement of some physical quantity of one of the particles 
affects the other at an instant. 
This has no classical counterpart, as shown by John Bell \cite{bell,bell_rmp,bell_local}. 
Nowadays, the Bell inequalities, tested in many experiments \cite{test1,test2,test4,test5,test6,test7,test8,test9,test10,test11,test12},
are analyzed in the context of the entanglement \cite{ent_rmp}.

Systems of entangled particles \cite{ent1,ent2} are potentially vialable to
quantum information \cite{quantinf}, teleportation \cite{wooters1,wooters2} or ultra-precise metrology \cite{giovannetti2004quantum,pezze2009entanglement}.
The definition of an entangled state is discriminative. Namely, it is based on the definition of
the separable (non-entangled) states as being a mixture of product states of $N$ particles \cite{sorensen2001many, peres, wer}, i.e., 
\begin{equation}\label{sep}
    \hat\varrho_{\rm sep}=\sum_i\,p_i\,\hat\varrho^{(1)}_i\otimes\ldots\otimes\hat\varrho^{(N)}_i.
\end{equation}
Here $p_i$'s are non-negative weights that add up to unity.
Entangled are all those states which cannot be represented in the form (\ref{sep}). A consequence of this indirect definition is that it is usually not straightforward to construct a criterion
or a measure of particle entanglement. Usually, one must find some physically measurable quantity $\mathcal A$, which for separable states is bounded (from above or below) by a critical value $\mathcal{A}_0$.
If some state gives the value of $\mathcal A$, which breaks this bound, then $\mathcal A$ is a criterion for particle entanglement. Trivially, such criterion detects entanglement of only those states,
which break the bound, therefore most $\mathcal A$'s are not universal.

Quantum interferometry provides a good example of such criterion. Namely, consider a two-mode state $\hat\varrho$ of $N$ particles passing through an interferometer. During the propagation,
a phase $\theta$ is imprinted between the modes. If the phase is estimated in a series of $m\gg1$ experiments, the precision of the estimation is limited by the Cram\'er-Rao lower bound 
\cite{holevo2011probabilistic, hells}:
\begin{equation}
  \Delta\theta\geqslant\frac1{\sqrt m}\frac1{\sqrt {F_Q}}.
\end{equation}
Here, $F_Q$ is the quantum Fisher information, which depends on $\hat\varrho$ and the type of applied interferometer \cite{braunstein1994statistical}. For all separable two-mode states $F_Q\leqslant N$
\cite{pezze2009entanglement}, therefore $\Delta\theta$ is limited by
the shot-noise. Nevertheless, there exist some ``usefully'' entangled states, for which $F_Q$ exceeds this critical value. Thus $F_Q=N$ plays the role of $\mathcal{A}_0$ mentioned above.

For atomic interferometers, the useful particle entanglement has been achieved by means of two-body
interactions present in ultra-cold systems. Usually, such correlations are associated with the spin-squeezing of a two-mode sample
\cite{kitagawa1993squeezed,wineland1994squeezed,esteve2008squeezing,appel2009mesoscopic,gross2010nonlinear,leroux2010orientation,riedel2010atom,chen2011conditional,berrada2013integrated,smerzi_ob}.
Alternatively, in a process which resembles the down-conversion,
the interactions drive the scattering of pairs of entangled atoms from a coherent source, such as a Bose-Einstein condensate (BEC)
\cite{collision_paris,lucke2011twin, twin_beam, cauchy_paris,twin_paris}.

In this work, we first show in Section \ref{sec_states} that the separable states of bosons formally resemble the classical states of electromagnetic field from Eq.~(\ref{sep}).
In Section \ref{sec_2nd} we derive the general expression for the second order correlation function of separable states. Rather then focusing on correlations between the position or momentum measurements,
we consider the most general measurement operators allowed by quantum mechanics.
In Section \ref{sec_csi} we argue that the violation of the Cauchy-Schwarz inequality (CSI) for the second order correlation function 
can be considered as a simple and useful criterion for entanglement of bosons.
We derive the CSI for generalized measurements and then focus on the correlations between the position measurements (Section \ref{sec_pos}).
We show, how this criterion can be extended to higher order correlations (Section \ref{higher}) and systems, where the number of particles is not fixed (Section \ref{sec_fl}). 
These Sections are an extension of our previous work \cite{cauchy}.
In Section \ref{sec_int} we further exploit the compact expression for the separable bosonic states to simplify the derivations of some well known interferometric criteria for particle entanglement.

\section{Separable state of $N$ bosons}
\label{sec_states}

In this Section, we relate the coherent states of light to separable states of $N$ bosons. 
Generally speaking, the non-classical electromagnetic field is that consisting of individual photons.
This statement can be quantified if the density matrix of the electromagnetic field is expressed using the P-representation \cite{sud,glaub}
\begin{equation}\label{em}
  \hat\varrho=\int\!\!\mathcal{D}\Phi\ket{\Phi}\!\!\bra{\Phi}\mathrm{P}(\Phi).
\end{equation}
Here, $\ket{\Phi}$ is a (possibly multi-mode) coherent state of light, defined by the relation
\begin{equation}
  \hat{\mathcal E}^{(+)}(\x)\ket\Phi=\Phi(\x)\ket\Phi.
\end{equation}
The operator $\hat{\mathcal E}^{(+)}(\x)$ implies the positive-frequency part of the electromagnetic field $\hat{\mathcal E}(\x)$, while
the symbol $\mathcal{D}\Phi$ in Eq.~(\ref{em}) is the integration measure over the set of complex fields $\Phi$.
In general, the state of light is called classical if the measurement outcomes can be explained in terms of classical electromagnetic waves.
It happens when the P-representation can be interpreted as a probability distribution, which occurs when $\mathcal{P}$ is normalized and
\begin{equation}\label{cond_em}
  \int\limits_{\mathcal V}\!\mathcal{D}\Phi\,\mathrm{P}(\Phi)F(\Phi)\geqslant0
\end{equation}
holds for {\it any} volume $\mathcal V$ and all non-negative functions $F(\Phi)$. In the opposite case, the field is not a mixture of coherent states and we call it quantum.

For a system of $N$ particles (distinguishable or not), a particle-separable state is defined by Eq.~(\ref{sep}). 
Any hermitian operator can be diagonalized, hence the $N$-body density matrix reads
\begin{equation}\label{rho}
  \hat\varrho =\sum_k \mathcal{P}_k \proj{\psi_k}.
\end{equation}
Here, $k$ is a composite index constructed from the index $i$ from Eq.~(\ref{sep}) and the indeces labeling the eigenstates of all $N$ one-body matrices. Similarly, the probabilities
$\mathcal{P}_k$ are products of $p_i$'s and the eigen-values of $\hat\varrho^{(j)}_i$. Each $\ket{\psi_k}$ is an $N$-body pure and separable state, which reads
\begin{equation}\label{sepket}
  \ket{\psi_k} = \ket{\phi_k^{(1)}} \otimes  \cdots \otimes \ket{\phi_k^{(N)}}.
\end{equation}
The upper index labels the particles and implies that the single-body eigen-vectors $\ket{\phi_k^{(n)}}$ and $\ket{\phi_k^{(n')}}$ might differ for $n\neq n'$.

We now restrict these general considerations to bosons. In such case the states $\ket{\psi_k}$ must be symmetrized with respect
to the particle interchange.
However, after the symmetrization of $\ket{\psi_k}$, the particles become entangled and consequently $\hat\varrho$ is not separable anymore. Clearly,
the only pure, symmetric and separable state must be a product of $N$ identical one-body states. Therefore, the upper index $n$ is no longer necessary and the state
from Eq.~(\ref{sepket}) transforms into
\begin{equation}\label{state_sym}
  \ket{\psi_k} = \ket{\phi_k}^{\otimes N}\equiv \ket{\phi_k;N}.
\end{equation}
Therefore, the set of density matrices of separable bosons is formed by mixtures of states (\ref{state_sym}) and reads
\begin{equation}\label{disc}
  \hat\varrho = \sum_k \mathcal{P}_k \proj{\phi_k;N}
\end{equation}
and when the labeling is done by a continuous parameter $\phi$
\begin{equation}\label{den}
  \hat\varrho=\int\!\!\mathcal{D}\phi\,\ket{\phi;N}\!\!\bra{\phi;N}\mathcal{P}(\phi).
\end{equation}
Here, symbol $\mathcal{D}\phi$ denotes the measure of integration over the set of states $\ket{\phi;N}$. 
The probability distribution $\mathcal{P}(\phi)$ is normalized and for all positive-defined functions $\mathcal{F}(\Phi)\geqslant0$ a relation
\begin{equation}\label{cond_bos}
  \int\limits_{\mathcal V}\!\mathcal{D}\Phi\,\mathcal{P}(\Phi)\mathcal{F}(\Phi)\geqslant0
\end{equation}
holds for any volume $\mathcal V$ \footnote{General form of separable states
of $N$ bosons has been already analyzed in different context, see for instance \cite{proof1,proof2}.}. 
The Equations (\ref{den}) and (\ref{cond_bos}) closely resemble Equations (\ref{em}) and (\ref{cond_em}) for electromagnetic fields, but while the former
pair is constructed with coherent states of light, which posses coherences between different particle-number states, the latter is valid for fixed $N$. 
If the density matrix cannot be written in the form (\ref{den}) with $\mathcal{P}(\phi)$ satisfying Eq.~(\ref{cond_bos}), the state is not separable so it is particle-entangled. 
In the following Sections we exploit this knowledge to derive some criteria for the separability of bosons.

\section{Second order correlation function}
\label{sec_2nd}

At this point, we formulate a criterion for the separability of a system of $N$ bosons, based on their second-order correlation function. We define this function using the 
positive-operator valued measures (POVMs) denoted by $\hat E(\xi)$.
These objects form the broadest set of measurement operators allowed by quantum mechanics \cite{quantinf,holevo2011probabilistic}. They are non-negative and form a complete basis, i.e.
\begin{equation}\label{norm}
  \hat E(\xi)\geqslant0\ ,\ \ \ \ \int\!\! \hat E(\xi)\,d\xi=\hat\mathds{1},
\end{equation}
where $\xi$ labels the detection outcomes. Most commonly, the measurements are modeled by the projection operators $\proj{\xi}$, which form a subset of POVMs, but for the moment we 
keep the discussion fully general.
We also assume that the measurements do not correlate particles, which means that the POVMs are single-body operators.

The second order correlation function is the rate of the coincidence of the results $\xi$ and $\xi'$ defined as
\begin{equation}
  G^{(2)}(\xi,\xi')=\sum_{i\neq j=1}^N\trc{\hat\varrho\,\hat E_i(\xi)\hat E_j(\xi')}.
\end{equation}
The labels $i$ and $j$ indicate that the operators $\hat E_i(\xi)$ and $\hat E_j(\xi')$ act on different particles, while the summation ensures that all coincidences contribute to the correlation function.
For indistinguishable particles these operators do not depend on the labels of the particles, therefore the indeces $i$ and $j$ can be dropped. The sum gives a combinatory prefactor
$\alpha_2=N(N-1)$ and upon using Eq.~(\ref{den}) we obtain
\begin{equation}\label{g2_sep}
  G^{(2)}(\xi,\xi')=\alpha_2\int\!\!\mathcal{D}\phi\,\mathcal{P}(\phi)\av{\hat E(\xi)}\av{\hat E(\xi')},
\end{equation}
where $\av{\hat E(\xi)}=\bra{\phi}\hat E(\xi)\ket\phi$. 
This is a general expression for the two-body correlation function for separable states of bosons. Any criterion for particle entanglement, which is based on the $G^{(2)}$
has to check if the second order correlation function can be written in the form of Eq.~(\ref{g2_sep}).

\section{Cauchy-Schwarz inequality as an entanglement criterion}
\label{sec_csi}

First, we derive a simple criterion for the particle entanglement using the CSI for the $G^{(2)}$ \cite{csi0, csi1, csi2, csi3, csi4, csi5, csi6, csi7}.
Consider two subsets $X_a$ and $X_b$ of the measurement outcomes and the second order correlation functions integrated over a pair of such regions
\begin{numparts}\label{int_corr22}
  \begin{eqnarray}
    \label{int_aa}&&\mathcal{G}^{(2)}_{aa}=\int\limits_{X_a}\!\! d\xi\!\!\int\limits_{X_a}\!\! d\xi'\,G^{(2)}(\xi,\xi')\label{integrals_aa},\\
    \label{int_bb}&&\mathcal{G}^{(2)}_{bb}=\int\limits_{X_b}\!\! d\xi\!\!\int\limits_{X_b}\!\! d\xi'\,G^{(2)}(\xi,\xi')\label{integrals_bb},\\
    \label{int_ab}&&\mathcal{G}^{(2)}_{ab}=\int\limits_{X_a}\!\! d\xi\!\!\int\limits_{X_b}\!\! d\xi'\,G^{(2)}(\xi,\xi')\label{integrals_ab}.
  \end{eqnarray}
\end{numparts}
The integrals (\ref{int_aa}) and (\ref{int_bb}) describe the effective local correlations within the sets $X_a$ and $X_b$ respectively. The last line (\ref{int_ab}) describes the cross-correlations
between the two sets. Note that for separable states, for which the $G^{(2)}$ is given by Eq.~(\ref{g2_sep}), we obtain 
\begin{equation}\label{integrated}
  \mathcal{G}^{(2)}_{ij}=\!\int\!\!\mathcal{D}\phi\,f_i(\phi)f_j(\phi),
\end{equation}
where $i$ and $j$ are either $a$ or $b$ and
\begin{equation}
  f_{i/j}(\phi)=\sqrt{\alpha_2\mathcal{P}(\phi)}\!\!\int\limits_{X_{i/j}}\!\!\!\! d\xi\,\av{\hat E(\xi)}.
\end{equation}
Since the functions $f_{a/b}(\phi)$ are non-negative, the CSI for integrals reads
\begin{equation}
\int\!\!\mathcal{D}\phi\,f_a(\phi)\,f_b(\phi)\leqslant\sqrt{\int\!\!\mathcal{D}\phi\, f^2_a(\phi)}\sqrt{\int\!\!\mathcal{D}\phi\, f^2_b(\phi)}\ \ .
\end{equation}
Therefore, we arrive at an inequality
\begin{equation}\label{csi}
  \mathcal{C}_2\equiv\frac{\mathcal{G}^{(2)}_{ab}}{\sqrt{\mathcal{G}^{(2)}_{aa}\mathcal{G}^{(2)}_{bb}}}\leqslant1,
\end{equation}
which is satisfied for all separable states of $N$ bosons\footnote{The volumes of the two regions in Eqs.~(\ref{int_aa})--(\ref{int_ab}) can be chosen arbitrarily small, and in fact the CSI also applies to correlations between two points.
However, from the experimental point of view it is usually advantageous to increase the signal by accumulating the data from substantial volumes.}. Any state breaking this bound must be particle--entangled.

\subsection{Special case -- position measurements}
\label{sec_pos}

We focus on an experimentally relevant case, when the POVMs represent the measurement of position. For the sake of clarity, we label the measurement outcomes with
$\x$ and $\x'$ rather then $\xi$ and $\xi'$. The POVM representing the idealized position measurement (with perfect accuracy and efficiency of the detector) is a projection onto the position state, i.e.
$\hat E(\x)=\proj{\x}$, which clearly satisfies conditions (\ref{norm}). 
The second order correlation function $G^{(2)}(\x,\x')$ measures the coincidence rate for detecting one particle at position $\x$ and the other at $\x'$. The correlation functions
in Eqs~(\ref{integrals_aa})--(\ref{integrals_ab}) are integrated over $a$ and $b$, which in this case denote two regions of space.

According to the CSI (Eq.~(\ref{csi})), for separable states of $N$ bosons, the cross-correlation between these regions cannot exceed the geometric
average of the local correlations. On the other hand, the enhanced cross-correlation might be signaled by the reduced fluctuations of the difference of the particle count between the two regions 
-- the so-called number squeezing. To link the CSI and the number squeezing
we express the second order correlation function using the bosonic field operator $\hat\Psi(\x)$, 
\begin{equation}\label{g2}
  G^{\,(2)}(\x,\x') = \left\langle \hat\Psi^\dagger(\x) \hat\Psi^\dagger(\x') \hat\Psi(\x')\hat\Psi(\x) \right\rangle.
\end{equation}
The atom number operator in each region is defined as
\begin{equation}\label{num_op}
  \hat n_{a/b}=\int\limits_{X_{a/b}}\!\!\! d\x\,\hat\Psi^\dagger(\x)\hat\Psi(\x).
\end{equation}
The fluctuations of the atom number difference operator $\hat n=\hat n_a-\hat n_b$ normalized to the shot-noise level $n_{\rm tot}=\av{\hat n_a}+\av{\hat n_b}$ give the number-squeezing parameter
\begin{equation}
  \eta^2=\frac{\av{(\Delta\hat n)^2}}{n_{\rm tot}}.
\end{equation}
Using Equations (\ref{g2}), (\ref{num_op}) and the integrated correlation functions (Eqs (\ref{integrals_aa})--(\ref{integrals_ab})) we obtain
\begin{equation}\label{ns}
  \eta^2=1+\frac{\mathcal{G}^{(2)}_{aa}+\mathcal{G}^{(2)}_{bb}-2\mathcal{G}^{(2)}_{ab}-\av{\hat n}^2}{n_{\rm tot}}.
\end{equation}
The system is number squeezed if $\eta^2<1$, which sets a constraint for the second part of the above equation, i.e.
\begin{equation}
  \eta^2<1\ \ \Leftrightarrow\ \ \mathcal{G}^{(2)}_{aa}+\mathcal{G}^{(2)}_{bb}-2\mathcal{G}^{(2)}_{ab}-\av{\hat n}^2<0.
\end{equation}
Although this condition resembles the CSI (\ref{csi}), the presence of the term $\av{\hat n}^2$ turns out to be crucial. To picture this, consider a separable pure state of $N$ particles
\begin{equation}
  \hat\varrho=\proj{\phi_0;N}
\end{equation}
divided into two unequal parts $a$ and $b$. For this coherent state the integrated second order correlation functions are related to the atom counts in the two regions, i.e.
$\mathcal{G}^{(2)}_{aa}=\av{\hat n_a}^2(1-\frac1N)$, $\mathcal{G}^{(2)}_{bb}=\av{\hat n_b}^2(1-\frac1N)$ and $\mathcal{G}^{(2)}_{ab}=\av{\hat n_a}\av{\hat n_b}(1-\frac1N)$. Therefore
\begin{equation}
  \eta^2=1-\frac{(\av{\hat n_a}-\av{\hat n_b})^2}{N^2}<1
\end{equation}
although the state is separable and does not violate the CSI. This example shows that the number-squeezing can be present in systems of bosons even in the absence of particle entanglement.

However, when there is symmetry between regions $a$ and $b$ such that $\langle\hat n\rangle=0$ and $\mathcal{G}^{(2)}_{aa}=\mathcal{G}^{(2)}_{bb}$ we obtain that
\begin{equation} \label{etasq}
  \eta^2=1-2\frac{(1-\mathcal{C}_2)\,\mathcal{G}^{(2)}_{aa}}{n_{\rm tot}}.
\end{equation}
In this case
\begin{equation}
  \eta^2<1\ \ \Leftrightarrow\ \ \mathcal{C}_2>1,
\end{equation}
so in such symmetric cases the number squeezing and CSI are strictly related. The violation of the CSI implies the number squeezing which signifies the particle entanglement.

\subsection{Extension to higher order correlations}
\label{higher}

The CSI discussed so far is based on the second order correlation function. Here we demonstrate that the arguments leading to Eq.~(\ref{csi}) can be extended to higher order correlations.
Consider an $2m$-th order correlation function between the measurement outcomes $\xi_1\ldots\xi_{2m}$. In analogy to expression for the $G^{(2)}$ from Eq.~(\ref{g2_sep}) we obtain that for separable states
\begin{equation}\label{gm}
  G^{\,(2m)}(\xi_1,\ldots,\xi_{2m})=\alpha_{2m}\int\!\!\mathcal{D}\phi\,\mathcal{P}(\phi)\,\av{\hat E(\xi_1)}\ldots\av{\hat E(\xi_{2m})},
\end{equation}
where $\alpha_{2m}=\frac{N!}{(N-2m)!}$. As in the case of the second order correlation function, we introduce two subsets of the outcomes $\xi$, integrate $m$ variables of
Eq.~(\ref{gm}) over  $X_a$ and other $m$ over $X_b$ obtaining
\begin{equation}\label{int_m}
  \mathcal{G}^{(2m)}_{ab}=\!\int\!\!\mathcal{D}\phi\,f^m_a(\phi)f^m_b(\phi).
\end{equation}
Similarly, $\mathcal{G}^{(2m)}_{aa}$ and $\mathcal{G}^{(2m)}_{bb}$ are the $2m$-th order correlation functions integrated $2m$-times over $a$ or $b$, respectively. The CSI applied to Eq.~(\ref{int_m}) gives
\begin{equation}\label{csi_m}
  \mathcal{C}_{2m}\equiv\frac{\mathcal{G}^{(2m)}_{ab}}{\sqrt{\mathcal{G}^{(2m)}_{aa}\mathcal{G}^{(2m)}_{bb}}}\leqslant1
\end{equation}
for any separable state of bosons.

\subsubsection{Application to twin-Fock state}

\begin{figure}[h]
  \includegraphics[clip, scale=0.34]{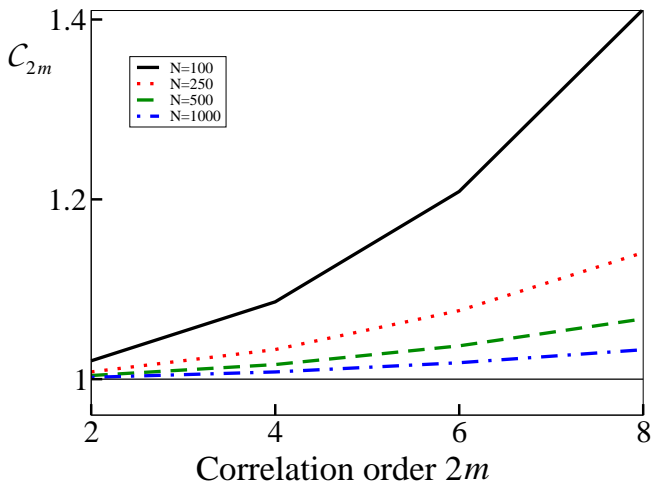}
  \caption{(Color online) The coefficient $\mathcal{C}_{2m}$ as a function of the correlation order for different numbers of particles in the twin-Fock state. Clearly, the deviation from the
    classical limit $\mathcal{C}_{2m}=1$ (shown here by a horizontal black line) increases with growing $m$.
  }\label{fig_2m}
\end{figure}

We argue that the strength of the CSI criterion increases when $m$ grows. To this end, consider a twin-Fock state of $N$ particles in two modes $a$ and $b$
\begin{equation}\label{tf}
  \ket\psi=\frac1{\left(\frac N2\right)!}\big(\hat b^\dagger\big)^{\frac N2}\big(\hat a^\dagger\big)^{\frac N2}|0,0\rangle=\left|\frac N2,\frac N2\right\rangle.
\end{equation}
This state is particle entangled, due to the indistinguishability of the constituent bosons. By identifying the modes with the two regions $X_a$ and $X_b$,
we analyze the CSI from Eq.~(\ref{csi_m}) as a function of $m$ for the twin-Fock state.
The local $\mathcal{G}^{(2m)}_{aa}$ correlation function is
\begin{equation}
  \mathcal{G}^{(2m)}_{aa}=\langle\big(\hat a^\dagger\big)^{2m}\big(\hat a\big)^{2m}\rangle=\frac{\left(\frac N2\right)!}{\left(\frac N2-2m\right)!}
\end{equation}
and symmetrically for $\mathcal{G}^{(2m)}_{bb}$. On the other hand, the cross correlation function is
\begin{equation}
      \mathcal{G}^{(2m)}_{ab}=\langle\big(\hat a^\dagger\hat b^\dagger\big)^{m}\big(\hat a\,\hat b\big)^{m}\rangle
      =\left(\frac{\left(\frac N2\right)!}{\left(\frac N2-m\right)!}\right)^2
\end{equation}
We substitute these results into Eq.~(\ref{csi_m}) and obtain
\begin{equation}\label{2m}
  \mathcal{C}_{2m}=\left(\frac N2\right)!\frac{\left(\frac N2-2m\right)!}{\left(\left(\frac N2-m\right)!\right)^2}.
\end{equation}
For given $N$, this is a growing function of the correlation order $2m$. 
In Fig.~\ref{fig_2m} we plot $\mathcal{C}_{2m}$ from Eq.~(\ref{2m}), $N=100$, 250, 500 and 1000 and the correlation order $2m=2,4,6,8$. 
For small values of the parameter $\epsilon = 2m/N$, Eq.~(\ref{2m}) can be approximated by
\begin{equation}
  \mathcal{C}_{2m} \approx \exp( \epsilon^2 N/2),
\end{equation}
which is in excellent agreement with the results shown in Fig.~\ref{fig_2m}.
Clearly, the deviation from the classical limit $\mathcal{C}_{2m}=1$ is higher when the ratio of $2m$ to $N$ increases, i.e. when the amount of information about the system
extracted from the high-order correlation function is large. 
This result might be of direct experimental relevance, since already a 6-th order correlation function among thermal atoms was measured \cite{dall2013ideal}. 

\subsection{Extension to systems with fluctuating $N$}
\label{sec_fl}

We will now demonstrate that the Cauchy-Schwarz criterion for particle entanglement can be successfully applied to systems where the number of particles differs from shot to shot. 
To account for these fluctuations
we impose the super-selection rule, which excludes coherences between states with different numbers of particles \cite{ssr_wick}. Consequently, a general separable state of bosons is a mixture
\begin{equation}\label{den_fl}
  \hat\varrho=\sum_{N=0}^\infty p_N\hat\varrho_N,
\end{equation}
where each $\hat\varrho_N$ is given by Eq.~(\ref{den}), while the statistical weights $p_N$ add up to unity. The second order correlation function calculated with the density matrix (\ref{den_fl}) is a
simple generalization of Eq.~(\ref{g2_sep}) and reads
\begin{equation}
  G^{(2)}(\xi,\xi')=\sum_{N=0}^\infty p_N\,\alpha_2(N)\!\!\int\!\!\mathcal{D}\phi\,\mathcal{P}_N(\phi)\av{\hat E(\xi)}\av{\hat E(\xi')}
\end{equation}
In this expression, $p_N$, $\alpha_2(N)$ and $\mathcal{P}_N(\phi)$ depends on $N$. On the other hand,
the average values of the single-particle operators $\hat E$ calculated with the one-particle states $\ket\phi$ do not depend on the number of particles. Therefore, it is reasonable to introduce
\begin{equation}
  \overline{\mathcal{P}}(\phi)=\sum_{N=0}^\infty p_N\,\alpha_2\mathcal{P}(\phi).
\end{equation}
This is an averaged $\mathcal P$ function, weighted with the coefficients $\alpha_2(N)$. One can introduce two regions $a$ and $b$ and the integrated second order correlation functions as in
Equations (\ref{integrals_aa})--(\ref{integrals_ab}). 
Those functions are given by Eq.~(\ref{integrated}) with $\alpha_2\mathcal{P}(\phi)$ replaced with $\overline{\mathcal{P}}(\phi)$. If $\mathcal{P}_N(\phi)$ for each
$N$ satisfies condition (\ref{cond_em}), so will $\overline{\mathcal{P}}(\phi)$, therefore, the CSI will be satisfied also by all separable states with fluctuating number of particles given by
Eq.~(\ref{den_fl}). Therefore, the violation of the CSI signals particle entanglement, just as when $N$ is fixed. 

However, there is one subtle difference between these two cases. Namely,
$\mathcal P_N$ can be partially negative for some $N$, which means that in this sector particles are entangled. Nevertheless the averaged $\overline{\mathcal{P}}(\phi)$ can still be
positive-defined, simply because the separable sectors of the density matrix might overshadow the entangled ones. Thus the CSI for systems with fluctuating number of particles can overlook
the particle entanglement present in some $N$-sector.
The above considerations apply also to higher order correlation functions and the related CSI formulated in Section \ref{higher}.

\section{Metrological entanglement criteria}
\label{sec_int}

Quantum metrology provides some useful criteria for entanglement of particles in two-mode systems. These criteria arise from
general considerations involving the upper bounds for the precision of the estimation of an unknown phase $\theta$,  imprinted on a system upon passing through a two-mode interferometer.
We show that the general expression for the density matrix of a collection of bosons from Eq.~(\ref{den}) vastly simplifies in the two-mode case.
Using this expression, we reformulate the well-known proofs that the quantum Fisher information and the spin-squeezing parameter are criteria for
particle entanglement.

\subsection{Two-mode separable states of bosons}

The two-mode states of bosons are described using the second quantization annihilation operators $\hat a$ and $\hat b$. We can define
the two-mode equivalent of the separable pure states defined in Eq.~(\ref{state_sym}). We put all $N$ bosons in the same quantum state which in the most general form reads
\begin{equation}\label{css}
  \ket{z,\varphi;N}=\sqrt{\frac1{N!}}\left(\sqrt z\,e^{i\varphi}\,\hat a^\dagger+\sqrt{1-z}\,\hat b^\dagger\right)^N\ket0.
\end{equation}
Here $z\in[0,1]$ measures the population imbalance between the two modes, while $\varphi\in[-\pi,\pi]$ is the relative phase.
The states $\ket{z,\varphi;N}$ are called the coherent spin states. The relations
\begin{numparts}
  \begin{eqnarray}
    &&\hat a\ket{z,\varphi;N}=\sqrt N\sqrt ze^{i\varphi}\ket{z,\varphi;N-1}\label{actiona}\\
    &&\hat b\ket{z,\varphi;N}=\sqrt N\sqrt{1-z}\ket{z,\varphi;N-1}\label{actionb}
  \end{eqnarray}
\end{numparts}
show the analogy with the states $\ket{\phi;N}$. Furthermore, the general two-mode state of bosons (in analogy to Eq.~(\ref{den})) is a mixture of different states (\ref{css}), i.e.
\begin{equation}\label{sep_2m}
  \hat\varrho=\int\limits_0^1\!\!dz\!\!\int\limits_{-\pi}^\pi\!\! d\varphi\,\mathcal{P}(z,\varphi)\ket{z,\varphi;N}\bra{z,\varphi;N}.
\end{equation}
Naturally, for $\hat\varrho$ to be particle-separable, $\mathcal{P}(z,\varphi)$ must posses the properties of the probability distribution as in Eq.~(\ref{cond_em}). 
With Eq.~(\ref{sep_2m}) at hand, we proceed to  the metrological criteria for particle entanglement.

\subsection{Quantum Fisher information for bosons}

When a collection of particles passes through an interferometer, a relative phase $\theta$ is imprinted between the two modes. The interferometer
can be usually represented by an evolution operator $\hat U=e^{-i\theta\hat h}$, where $\hat h$ is the generator of the unitary transformation.
Subsequently, some measurement is performed at the output to deduce the value of
the parameter. Usually, the sequence of measurements is repeated $m\gg1$ times, to obtain the average value of the estimated parameter and the associated fluctuations $\Delta\theta$. If the estimator is
unbiased, i.e. it has the desired property that this average tends to the true value of the parameter, then according to the Cram\'{e}r-Rao lower bound \cite{holevo2011probabilistic, hells} 
the fluctuations are constrained by
\begin{equation}
  \Delta\theta\geqslant\frac1{\sqrt m}\frac1{\sqrt{F_Q[\hat\varrho]}}.
\end{equation}
Here $F_Q$ is the quantum Fisher information (QFI), which depends on both the density matrix $\hat\varrho$ and the interferometric transformation \cite{braunstein1994statistical}. 
The expression for $F_Q$ is in general rather complicated but for pure states $\hat\varrho$ it reduces to
\begin{equation}\label{ineq2_fq}
  F_Q[\hat\varrho]=4\langle\big(\Delta\hat h\big)^2\rangle,
\end{equation}
where $\Delta\hat h=\hat h-\langle\hat h\rangle$. The average values are calculated in the input state $\hat\varrho$.
When $\hat h$ is a one-body operator, so the interferometer does not correlate the particles, then
\begin{equation}\label{ineq_fq}
  F_Q[\hat\varrho]\leqslant N
\end{equation}
for all separable two-mode states of $N$ particles. Therefore $F_Q>N$ signals the entanglement between the particles. The original proof of this statement dealt with general separable two-mode states
which can be written in the form of Eq.~(\ref{sep}) \cite{pezze2009entanglement}. 

We show how this derivation simplifies for two-mode states of bosons.
Note that the two-mode interferometric transformations are generated by the angular momentum operators
\begin{numparts}
  \begin{eqnarray}
    &&\hat J_x=\frac12\left(\hat a^\dagger\hat b+\hat b^\dagger\hat a\right)\\
    &&\hat J_y=\frac1{2i}\left(\hat a^\dagger\hat b-\hat b^\dagger\hat a\right)\\
    &&\hat J_z=\frac12\left(\hat a^\dagger\hat a-\hat b^\dagger\hat b\right),
  \end{eqnarray}
\end{numparts}
which form the closed Lie algebra of the angular momentum. The generic interferometric transformation $\hat U=e^{-i\theta\hat J_{\vec n}}$, where $\hat J_{\vec n}=\vec n\cdot\vec{\hat J}$ is a scalar product
of a unit vector and $\vec{\hat J}=(\hat J_x,\hat J_y,\hat J_z)^T$.
Note that $\hat J_{\vec n}$ for any $\vec n$ can be obtained by rotating the $\hat J_z$ operator, namely
\begin{equation}
    \hat J_{\vec n}=e^{i\vec v_{\vec n}\vec{\hat J}}\hat J_ze^{-i\vec v_{\vec n}\vec{\hat J}},
\end{equation}
where $\vec v_{\vec n}$ is some vector oriented in a direction depending on $\vec n$. In the evaluation of Eq.~(\ref{ineq2_fq}), these two external rotations $e^{\pm i\vec v_{\vec n}\vec{\hat J}}$
can be moved so that they
transform the density matrix rather than the operator $\hat J_z$. Since these operations do not entangle particles, the transformed density matrix of a separable state will still have the form of 
Eq.~(\ref{sep_2m}). This means, that it is sufficient to show that  $F_Q\leqslant N$
for the $\hat J_z$ transformation and any separable two-mode state. According to the above argument, this inequality will then hold for any $\hat J_{\vec n}$. 

One important property of the QFI is its convexity, namely
\begin{equation}\label{con0}
  F_Q\left[\sum_i p_i\hat\varrho_i\right]\leqslant \sum_i p_iF_Q[\hat\varrho_i].
\end{equation}
This property, applied to states from Eq.~(\ref{sep_2m}) gives
\begin{equation}\label{convex}
  F_Q\left[\hat\varrho\right]\leqslant\int\limits_0^1\!\!dz\!\!\int\limits_{-\pi}^\pi\!\! d\varphi\,\mathcal{P}(z,\varphi)F_Q\left[\ket{z,\varphi;N}\bra{z,\varphi;N}\right].
\end{equation}
Now the QFI within the integral is calculated for a the pure state $\hat\varrho_0=\ket{z,\varphi;N}\bra{z,\varphi;N}$ and thus it is simply given by four times the variance of the $\hat J_z$ operator.
With help of the relations (\ref{actiona}) and (\ref{actionb}) we arrive at
\begin{equation}
  F_Q\left[\hat\varrho_0\right]=4\langle\big(\Delta\hat J_z\big)^2\rangle=4Nz(1-z)\leqslant N
\end{equation}
for $z\in[0,1]$. Using the property from Eq.~(\ref{convex}) we get
\begin{equation}\label{res_fix}
  F_Q\left[\hat\varrho\right]\leqslant N
\end{equation}
for all separable states. We conclude that $F_Q>N$ is a criterion for particle entanglement.

\subsubsection{Fluctuating $N$}

The above arguments can be extended to situations when the number of particles fluctuates in the two-mode system \cite{fluct_smerzi}. 
In such case in the absence of coherence between states with different numbers of particles the separable two-mode state of bosons is given by Eq.~(\ref{den_fl}),
where $\hat\varrho_N$ is given by Eq.~(\ref{sep_2m}). For each sector of fixed $N$, we can repeat the derivation leading to
Eq.~(\ref{res_fix}) and using the convexity property (\ref{con0}), we obtain 
\begin{equation}
  F_Q\left[\hat\varrho\right]\leqslant\sum_{N=0}^\infty p_N N=\langle N\rangle.
\end{equation}
Thus $F_Q>\langle N\rangle$ singals the entanglement between the particles.

\subsection{Spin-squeezing for bosons}

The QFI is a powerful criterion for particle entanglement, which is useful for metrological purposes. However, it has one major drawback -- it is hard to determine in the experiment.
Quantum interferometry provides much simpler entanglement criteria, which do not require the full knowledge of the density matrix $\hat\varrho$. Among these is the spin-squeezing parameter
\cite{kitagawa1993squeezed,wineland1994squeezed}
\begin{equation}\label{xi}
  \xi_s^2=N\frac{\av{\big(\Delta\hat J_z\big)^2}}{\av{\hat J_x}^2+\av{\hat J_y}^2}.
\end{equation}
This parameter is bounded for two-mode separable states by $\xi_s^2\geqslant1$. 
Therefore $\xi_s^2<1$ implies the particle entanglement. As in the previous Section, we can show that the general proof of this fact
presented in \cite{sorensen2001many} simplifies for separable states of indistinguishable bosons.

Note that with help of relations (\ref{actiona}) and (\ref{actionb}) we obtain that
\begin{numparts}
  \begin{eqnarray}
    &&\av{\hat J_x}=N\int dz\int d\varphi P(z,\varphi)\sqrt{z(1-z)}\cos\varphi\label{moma}\\
    &&\av{\hat J_y}=N\int dz\int d\varphi P(z,\varphi)\sqrt{z(1-z)}\sin\varphi\label{momab}\\
    &&\av{\big(\Delta\hat J_z\big)^2}=\frac N4+N(N-1)\int dzP(z)\left(z-\frac12\right)^2+\nonumber\\
    &&\phantom{\av{\big(\Delta\hat J_z\big)^2}=}-N^2\left[\int dzP(z)\left(z-\frac12\right)\right]^2.\label{var_z}\label{momc}
  \end{eqnarray}
\end{numparts}
We now apply the Cauchy-Schwarz inequality for integrals
\begin{numparts}
  \begin{eqnarray}
    &&\av{\hat J_x}^2\leqslant N^2\!\!\int\!\! dz P(z)z(1-z)\int\!\!d\varphi P(\varphi)\cos^2\!\varphi\\
    &&\av{\hat J_y}^2\leqslant N^2\!\!\int\!\! dz P(z)z(1-z)\int\!\!d\varphi P(\varphi)\sin^2\!\varphi.
  \end{eqnarray}
\end{numparts}
Therefore, the denominator of Eq.~(\ref{xi}) can be bounded as follows
\begin{equation}
  \av{\hat J_x}^2+\av{\hat J_y}^2\leqslant N^2\int dz P(z)z(1-z).
\end{equation}
The spin-squeezing parameter is thus not smaller then
\begin{equation}
  \xi_s^2\geqslant\frac{\av{\big(\Delta\hat J_z\big)^2}}{N\int dz P(z)z(1-z)},
\end{equation}
where the nominator is given by Eq.~(\ref{var_z}). Our aim is to show that $\xi_s^2\geqslant1$ or equivalently
\begin{equation}
  \av{\big(\Delta\hat J_z\big)^2}\geqslant N\int dz P(z)z(1-z).
\end{equation}
Using (\ref{var_z}), the above inequality gives
\begin{equation}\label{sq_res}
  \int dzP(z)\left(z-\frac12\right)^2-\left[\int dzP(z)\left(z-\frac12\right)\right]^2\geqslant0,
\end{equation}
which is always true, since it is a variance of the variable $z-\frac12$. With this we have proven that $\xi_s^2\geqslant1$ for all separable two-mode states of bosons.

\subsubsection{Fluctuating $N$}

When the number of particles is not fixed, the moments of the angular momentum operators are calculated using the two mode state (\ref{den_fl}). 
As a result, the Equations (\ref{moma})-(\ref{momc}) are transformed into
\begin{numparts}
  \begin{eqnarray}
    &&\label{var_x2}\av{\hat J_x}=\langle N\rangle\int dz\int d\varphi P(z,\varphi)\sqrt{z(1-z)}\cos\varphi\\
    &&\label{var_y2}\av{\hat J_y}=\langle N\rangle\int dz\int d\varphi P(z,\varphi)\sqrt{z(1-z)}\sin\varphi\\
    &&\av{\big(\Delta\hat J_z\big)^2}=\frac {\langle N\rangle}4+\left(\langle N^2\rangle-\langle N\rangle^2\right)\int dzP(z)\left(z-\frac12\right)^2\nonumber\\
    &&\phantom{\av{\big(\Delta\hat J_z\big)^2}=}-\langle N\rangle^2\left[\int dzP(z)\left(z-\frac12\right)\right]^2.\label{var_z2}
  \end{eqnarray}
\end{numparts}
The Cauchy-Schwarz inequality can again be applied to Equations (\ref{var_x2}) and (\ref{var_y2}) giving
\begin{equation}
  \av{\hat J_x}^2+\av{\hat J_y}^2\leqslant \langle N\rangle^2\int dz P(z)z(1-z).
\end{equation}
Upon performing manipulations similar to those leading to Eq.~(\ref{sq_res}) but with $N$ replaced with $\langle N\rangle$ and using $\langle N^2\rangle\geqslant\langle N\rangle^2$
obtain that
\begin{equation}\label{xi2}
  \langle N\rangle \frac{\av{\big(\Delta\hat J_z\big)^2}}{\av{\hat J_x}^2+\av{\hat J_y}^2}\geqslant1
\end{equation}
for all separable two-mode states of bosons with the fluctuating number of particles. Therefore we proved that the properly defined spin-squeezing parameter is also an entanglement criterion in the
non fixed-$N$ case.

\section{Conclusions}
We argued that ample experience of quantum optics, and in particular its methods to tackle the notion of classical, semiclassical and quantum, provide tools to describe quantum correlations 
in many body systems. We considered a general POVM -- a measurement operator allowed by quantum mechanics -- and focused on the correlation functions of the outcomes of such measurement. 
We proved that when system is in a separable state, the correlations between the measurement outcomes 
satisfy the Cauchy-Schwarz inequality, and hence we drew the conclusion that when it is violated it can be treated as a criterion for 
non-classicality of bosonic states.  Our results are general, they apply to any collection of bosons, even when their number fluctuates from shot to shot. 
The only constraint is that the super selection rule applies, forbidding coherences between different number states. 
Finally we take advantage of the simple form of the separable state of bosons to express known interferometric criteria for particle entanglement, such as the QFI or the spins-squeezing in a simple manner.

\section{Acknowledgments}

T. W. acknowledges the Foundation for Polish Science International Ph.D. Projects Programme co-financed by the EU European Regional Development Fund.
T. W., P. Sz. and J. Ch. were supported by the National Science Center grant no. DEC-2011/03/D/ST2/00200.

\section*{References}

\providecommand{\newblock}{}


\begin{thebibliography}{10}
\expandafter\ifx\csname url\endcsname\relax
  \def\url#1{{\tt #1}}\fi
\expandafter\ifx\csname urlprefix\endcsname\relax\def\urlprefix{URL }\fi
\providecommand{\eprint}[2][]{\url{#2}}

\bibitem{epr}
Einstein A, Podolsky B and Rosen N {\em Phys. Rev.\/} {\bf 47}(10) 777

\bibitem{bell}
Bell J~S 1964 {\em Physics\/} {\bf 1} 195

\bibitem{bell_rmp}
Bell J~S 1966 {\em Rev. Mod. Phys.\/} {\bf 38}(3) 447--452

\bibitem{bell_local}
Brunner N, Cavalcanti D, Pironio S, Scarani V and Wehner S 2014 {\em Rev. Mod.
  Phys.\/} {\bf 86}(2) 419--478

\bibitem{test1}
Freedman S~J and Clauser J~F 1972 {\em Phys. Rev. Lett.\/} {\bf 28}(14)
  938--941

\bibitem{test2}
Aspect A, Grangier P and Roger G 1981 {\em Phys. Rev. Lett.\/} {\bf 47}(7)
  460--463

\bibitem{test4}
Tittel W, Brendel J, Gisin B, Herzog T, Zbinden H and Gisin N 1998 {\em Phys.
  Rev. A\/} {\bf 57}(5) 3229--3232

\bibitem{test5}
Tittel W, Brendel J, Zbinden H and Gisin N 1998 {\em Phys. Rev. Lett.\/} {\bf
  81}(17) 3563--3566

\bibitem{test6}
Weihs G, Jennewein T, Simon C, Weinfurter H and Zeilinger A 1998 {\em Phys.
  Rev. Lett.\/} {\bf 81}(23) 5039--5043

\bibitem{test7}
{Pan} J~W, {Bouwmeester} D, {Daniell} M, {Weinfurter} H and {Zeilinger} A 2000
  {\em Nature\/} {\bf 403} 515--519

\bibitem{test8}
{Kielpinski} D, {Meyer} V, {Sackett} C~A, {Itano} W~M, {Monroe} C and
  {Wineland} D~J 2001 {\em Nature\/} {\bf 409} 791--794

\bibitem{test9}
{Gr{\"o}blacher} S, {Paterek} T, {Kaltenbaek} R, {Brukner} {\v C},
  {{\.Z}ukowski} M, {Aspelmeyer} M and {Zeilinger} A 2007 {\em Nature\/} {\bf
  446} 871--875

\bibitem{test10}
Salart D, Baas A, van Houwelingen J~A~W, Gisin N and Zbinden H 2008 {\em Phys.
  Rev. Lett.\/} {\bf 100}(22) 220404

\bibitem{test11}
{Ansmann} M, {Wang} H, {Bialczak} R~C, {Hofheinz} M, {Lucero} E, {Neeley} M,
  {O'Connell} A~D, {Sank} D, {Weides} M, {Wenner} J, {Cleland} A~N and
  {Martinis} J~M 2009 {\em Nature\/} {\bf 461} 504--506

\bibitem{test12}
{Giustina} M, {Mech} A, {Ramelow} S, {Wittmann} B, {Kofler} J, {Beyer} J,
  {Lita} A, {Calkins} B, {Gerrits} T, {Nam} S~W, {Ursin} R and {Zeilinger} A
  2013 {\em Nature\/} {\bf 497} 227--230

\bibitem{ent_rmp}
Horodecki R, Horodecki P, Horodecki M and Horodecki K 2009 {\em Rev. Mod.
  Phys.\/} {\bf 81}(2) 865--942

\bibitem{ent1}
Bengtsson I and \.Zyczkowski K 2006 {\em Geometry of Quantum States. An
  Introduction to Quantum Entanglement\/} (Cambridge University Press)

\bibitem{ent2}
Horodecki R, Horodecki P, Horodecki M and Horodecki K 2009 {\em Rev. Mod.
  Phys.\/} {\bf 81}(2) 865

\bibitem{quantinf}
Nielsen M~A and Chuang I~L 2000 {\em Quantum computation and quantum
  information\/} (Cambridge University Press)

\bibitem{wooters1}
Bennett C~H, Brassard G, Cr\'epeau C, Jozsa R, Peres A and Wootters W~K 1993
  {\em Phys. Rev. Lett.\/} {\bf 70}(13) 1895

\bibitem{wooters2}
Bennett C~H, Brassard G, Popescu S, Schumacher B, Smolin J~A and Wootters W~K
  1996 {\em Phys. Rev. Lett.\/} {\bf 76} 722

\bibitem{giovannetti2004quantum}
Giovannetti V, Lloyd S and Maccone L 2004 {\em Science\/} {\bf 306} 1330--1336

\bibitem{pezze2009entanglement}
Pezz{\'e} L and Smerzi A 2009 {\em Phys. Rev. Lett.\/} {\bf 102} 100401

\bibitem{sorensen2001many}
S{\o}rensen A, Duan L~M, Cirac J and Zoller P 2001 {\em Nature\/} {\bf 409}
  63--66

\bibitem{peres}
Peres A 1996 {\em Phys. Rev. Lett.\/} {\bf 77} 1413--1415

\bibitem{wer}
Werner R~F 1989 {\em Phys. Rev. A\/} {\bf 40} 4277--4281

\bibitem{holevo2011probabilistic}
Holevo A 2011 {\em Probabilistic and Statistical Aspects of Quantum Theory\/}
  (Publications of Scuola Normale Superiore)

\bibitem{hells}
Helstrom C~W 1969 {\em Journal of Statistical Physics\/} {\bf 1} 231--252

\bibitem{braunstein1994statistical}
Braunstein S~L and Caves C~M 1994 {\em Phys. Rev. Lett.\/} {\bf 72} 3439--3443

\bibitem{kitagawa1993squeezed}
Kitagawa M and Ueda M 1993 {\em Phys. Rev. A\/} {\bf 47} 5138--5143

\bibitem{wineland1994squeezed}
Wineland D, Bollinger J, Itano W and Heinzen D 1994 {\em Phys. Rev. A\/} {\bf
  50} 67

\bibitem{esteve2008squeezing}
Esteve J, Gross C, Weller A, Giovanazzi S and Oberthaler M 2008 {\em Nature\/}
  {\bf 455} 1216--1219

\bibitem{appel2009mesoscopic}
Appel J, Windpassinger P~J, Oblak D, Hoff U~B, Kj{\ae}rgaard N and Polzik E~S
  2009 {\em PNAS\/} {\bf 106} 10960--10965

\bibitem{gross2010nonlinear}
Gross C, Zibold T, Nicklas E, Esteve J and Oberthaler M~K 2010 {\em Nature\/}
  {\bf 464} 1165--1169

\bibitem{leroux2010orientation}
Leroux I~D, Schleier-Smith M~H and Vuleti{\'c} V 2010 {\em Phys. Rev. Lett.\/}
  {\bf 104} 250801

\bibitem{riedel2010atom}
Riedel M~F, B{\"o}hi P, Li Y, H{\"a}nsch T~W, Sinatra A and Treutlein P 2010
  {\em Nature\/} {\bf 464} 1170--1173

\bibitem{chen2011conditional}
Chen Z, Bohnet J~G, Sankar S~R, Dai J and Thompson J~K 2011 {\em Phys. Rev.
  Lett.\/} {\bf 106} 133601

\bibitem{berrada2013integrated}
Berrada T, van Frank S, B{\"u}cker R, Schumm T, Schaff J~F and Schmiedmayer J
  2013 {\em Nat. Commun.\/} {\bf 4}

\bibitem{smerzi_ob}
Strobel H, Muessel W, Linnemann D, Zibold T, Hume D~B, Pezz\'e L, Smerzi A and
  Oberthaler M~K 2014 {\em Science\/} {\bf 345} 424--427

\bibitem{collision_paris}
Perrin A, Chang H, Krachmalnicoff V, Schellekens M, Boiron D, Aspect A and
  Westbrook C~I 2007 {\em Phys. Rev. Lett.\/} {\bf 99}(15) 150405

\bibitem{lucke2011twin}
L{\"u}cke B, Scherer M, Kruse J, Pezz{\'e} L, Deuretzbacher F, Hyllus P, Peise
  J, Ertmer W, Arlt J, Santos L {\em et~al.\/} 2011 {\em Science\/} {\bf 334}
  773--776

\bibitem{twin_beam}
B\"ucker R, Grond J, Manz S, Berrada T, Betz T, Koller C, Hohenester U, Schumm
  T, Perrin A and Schmiedmayer J 2011 {\em Nat. Phys.\/} {\bf 7} 608

\bibitem{cauchy_paris}
Kheruntsyan K~V, Jaskula J~C, Deuar P, Bonneau M, Partridge G~B, Ruaudel J,
  Lopes R, Boiron D and Westbrook C~I 2012 {\em Phys. Rev. Lett.\/} {\bf
  108}(26) 260401

\bibitem{twin_paris}
Bonneau M, Ruaudel J, Lopes R, Jaskula J~C, Aspect A, Boiron D and Westbrook
  C~I 2013 {\em Phys. Rev. A\/} {\bf 87}(6) 061603

\bibitem{cauchy}
Wasak T, Sza\'{n}kowski P, Zi\'{n} P, Trippenbach M and Chwede\'{n}czuk J 2014
  {\em Phys. Rev. A\/} {\bf 90} 033616

\bibitem{sud}
Sudarshan E~C~G 1963 {\em Phys. Rev. Lett.\/} {\bf 10}(7) 277

\bibitem{glaub}
Glauber R~J 1963 {\em Phys. Rev.\/} {\bf 131}(6) 2766

\bibitem{proof1}
Ichikawa T, Sasaki T, Tsutsui I and Yonezawa N 2008 {\em Phys. Rev. A\/} {\bf
  78} 052105

\bibitem{proof2}
Wei T~C 2010 {\em Phys. Rev. A\/} {\bf 81} 062313

\bibitem{csi0}
Tura J, Augusiak R, Sainz A~B, Vértesi T, Lewenstein M and Acín A 2014  {\bf
  344} 1256

\bibitem{csi1}
Clauser J~F 1976 {\em Phys. Rev. Lett.\/} {\bf 36} 1223

\bibitem{csi2}
Shchukin E and Vogel W 2005 {\em Phys. Rev. Lett.\/} {\bf 95} 230502

\bibitem{csi3}
Hillery M and Zubairy M~S 2006 {\em Phys. Rev. Lett.\/} {\bf 96} 050503

\bibitem{csi4}
Miranowicz A, Bartkowiak M, Wang X, Liu Y~x and Nori F 2010 {\em Phys. Rev.
  A\/} {\bf 82} 013824

\bibitem{csi5}
de~Nova J~R~M, Sols F and Zapata I 2014 {\em Phys. Rev. A\/} {\bf 89} 043808

\bibitem{csi6}
Busch X, Carusotto I and Parentani R 2014 {\em Phys. Rev. A\/} {\bf 89} 043819

\bibitem{csi7}
Englert B~G and W\'odkiewicz K 2002 {\em Phys. Rev. A\/} {\bf 65} 054303

\bibitem{dall2013ideal}
Dall R, Manning A, Hodgman S, RuGway W, Kheruntsyan K and Truscott A 2013 {\em
  Nature Physics\/} {\bf 9} 341--344

\bibitem{ssr_wick}
Wick G~C, Wightman A~S and Wigner E~P 1952 {\em Phys. Rev.\/} {\bf 88} 101--105

\bibitem{fluct_smerzi}
Hyllus P, Pezz\'e L and Smerzi A 2010 {\em Phys. Rev. Lett.\/} {\bf 105}(12)
  120501

\end{thebibliography}


\end{document}